\documentclass[showpacs,preprintnumbers,amsmath,amssymb,titlepage]{revtex4}

\usepackage{epsfig}

\begin{document}

\title{Mean-field theory of orientational ordering in rigid rotor
  models with identical atoms: spin conversion and thermal
  equilibration  \\ \vspace{1cm} Short title: Mean-field theory of coupled rotors}

\author{ Bal\'azs Het\'enyi }

\affiliation{Institut f\"ur Theoretische Physik, Technische Universit\"at Graz, \\ Petersgasse 16, Graz A-8010, Austria \\ and \\
Mathematisches Institut, \\ Ludwig Maximilians Universit\"at, \\ Theresienstrasse 39, M\"unchen 80333, Germany
}


\begin{abstract}
  In coupled rotor models which describe identical rotating nuclei the
  nuclear spin states restrict the possible angular momenta of each
  molecule.  There are two mean-field approaches to determining the
  orientational phase diagrams in such systems.  In one the nuclear
  spin conversion times are assumed to be instantaneous in the other
  infinite.  In this paper the intermediate case, when the spin
  conversion times are significantly slower than those of rotational
  time scales, but are not infinite on the time-scale of the
  experiment, is investigated.  Via incorporation of the
  configurational degeneracy it is shown that in the thermodynamic
  limit the mean-field approach in the intermediate case is identical
  to the instantaneous spin conversion time approximation.  The total
  entropy can be split into configurational and rotational terms.  The
  mean-field phase diagram of a model of coupled rotors of three-fold
  symmetry is also calculated in the two approximations.  It is shown
  that the configurational entropy has a maximum as a function of
  temperature which shifts to lower temperatures with increasing
  order.
\end{abstract}

\maketitle

\section{Introduction}
\label{sec:intro}

Molecules in crystal phases often exhibit quantum orientational ordering.  A
well-known example is solid hydrogen~\cite{Silvera80,Mao94}, where the phase
diagrams are strongly dependent on the nuclear spin of the molecules.  At low
pressure the free rotor angular momentum quantum numbers can still be
considered good.  The symmetry requirement of the molecular wavefunction
imposes constraints on the allowed angular momentum quantum numbers for a
molecule of a given nuclear spin configuration.  In the case of H$_2$
molecules in the triplet(singlet) nuclear spin state are coupled to odd(even)
angular momentum levels.  In D$_2$ the situation is similar, the
degeneracies, however, are different: even(odd) angular momentum states are
sixfold(threefold) degenerate~\cite{deg}.  Nuclear spin has important effects
on orientational ordering.  In H$_2$ the pure para system orders only at high
pressures, $\approx 100 GPa $ (see Fig. 1b of Ref.  \cite{Cui95}), whereas
the ground state of ortho-H$_2$ is ordered at finite pressure.  When both
even and odd angular momenta are accessible, as is the case in the
heteronuclear species
HD~\cite{Freiman91,Brodyanskii93,Freiman09,Moshary96,Shin09}, or when the
ortho-para distribution in H$_2$ is thermally equilibrated
~\cite{Hetenyi05,Freiman05,Goncharenko05}, then the phase diagram is
reentrant.  Very similar patterns of phase diagrams are also
found~\cite{Mueser99,Hetenyi99,Hetenyi05b} in the quantum anisotropic planar
rotor model~\cite{Martonak97}, in which uniaxial rotors corresponding to
diatomic molecules are coupled, and in similar models of coupled uniaxial
rotors~\cite{Parmenter67,Simanek85,Simkin91}.  Other examples of systems
where models of coupled rotors have played a role in elucidating the physics
are inclusion compounds~\cite{Sloan90,Hubbard81,Lunine87,Loveday01}
containing ammonia, Hoffmann
clathrates~\cite{Buettner94,Iwamoto74,Sobolev04}, and crystals containing
methyl groups~\cite{Press81,Prager97}.  Regarding these systems the issue of
how nuclear spin effects ordering has not been thoroughly studied up to now.

The mean-field approach has been used to understand the orientational
ordering in coupled rotor systems.  In solid H$_2$ the mean-field
approach predicts the correct qualitative behaviour for sytems of pure
para or ortho species, and of HD, where nuclear spin is not coupled to
angular momentum, hence the angular momentum distribution is allowed
to thermally equilibrate instantaneously.  In solid H$_2$ it is known
that the spin conversion times are long compared to rotational
time-scales, hence thermal equilibrium is not instantenous on
rotational time-scales at low pressures.  This is also thought to be
the case in ammonia or methyl containing crystals~\cite{Wurger90}.  In
ammonia or methyl containing crystals the effects of nuclear spin on
orientational ordering have not been investigated.

Mean-field studies of hydrogen assume that the nuclear spin conversion
times are either infinite or instantaneous on rotational
time-scales~\cite{Freiman05}.  In this work the intermediate case,
where nuclear spin relaxation is long on the time-scale of rotation
but allowed to equilibrate on the experimental time-scale is
investigated.  It is shown that in the thermodynamic limit the
mean-field results for such a case are equivalent to the zero
relaxation time approximation.  The total entopy is also shown to
consist of two contributions, one configurational and one rotational.

The mean-field phase diagram of a model of three-fold symmetric
coupled rotors is also presented.  Here the nuclear spin states can be
of three types which couple to three different manifolds of angular
momentum ($A$, $E_a$, and $E_b$), of which two ($E_a$ and $E_b$) are
equivalent, hence display the same ordering properties.  The thermal
mixture displays a reentrant phase transition (as predicted
prior~\cite{private}). The reentrant phase transition is known to be
accompanied by an entropy anomaly~\cite{Freiman05}.  At low
temperatures the entropy in the ordered state is larger than in the
disordered state.  In this study the two components of the entropy
(configurational and rotational) are calculated and it is shown that
both display similar anomalous behaviour in the reentrant region.  The
maximum of the configurational entropy shifts to smaller values of
temperature as the order parameter is increased.

This paper is organized as follows.  In section \ref{sec:mean-field
  theory} mean-field theory of coupled rotors is discussed in the
context of nuclear spin.  Subsequently the model studied here, that of
coupled three-fold symmetric uniaxial rotors, is discussed.  In
section \ref{sec:results} the results are presented and in section
\ref{sec:conclusions} conclusions are drawn.

\section{Mean-field theory for systems with nuclear spin}
\label{sec:mean-field theory}

In the following a model of coupled rigid rotors in which the nuclear
spin configurations are coupled to angular momentum states as is the
situation in systems with identical rotating atoms is considered.  Let
the model consist of $N$ rotors whose centers of mass are fixed.  Let
there be $\sigma$ different manifold of angular momentum states
coupled to a particular manifold of spin states of degeneracy
$G_{\sigma}$.  For example in solid hydrogen (H$_2$) $\sigma=2$, as
there is the ortho and para variety, and the degeneracies are $G_1=1$
(singlet) and $G_2=3$ (triplet).

To calculate the phase diagram a Hamiltonian of the following form is
considered
\begin{equation}
  H = -B \sum_i \frac{\partial^2}{\partial \phi^2} - J \sum_{\langle i,j \rangle} \mbox{cos}(n\phi_i) \mbox{cos}(n\phi_j),
\label{eqn:Hml_tot}
\end{equation}
where $B$ denotes the rotational constant, and $J$ denotes the coupling
between rotors, $\phi_i$ denotes the angular coordinate of rotor at site $i$,
and $\langle i,j \rangle$ denotes summation over nearest neighbors.  In this
work we apply a mean-field type approximation to the Hamiltonian in
Eq. (\ref{eqn:Hml_tot}), e.g.
\begin{equation}
H_{MF} = -B \frac{\partial^2}{\partial \phi^2} - J \Gamma \mbox{cos}(n\phi)
 + J \frac{\Gamma^2}{2}.
\label{eqn:Hml}
\end{equation}
This is a model of coupled uniaxial rotors, but the conclusions in
this section are valid for rotors in any number of dimensions.  The
model (Eq. (\ref{eqn:Hml_tot})) includes a simplified potential
sensitive only to orientation and the molecular symmetry.  While the
crystal structure and phonons may also effect orientational ordering,
the point of view taken here is that orientation and molecular
symmetry are the most important.  This point of view is corroborated
by the mean-field results of previous studies on
hydrogen~\cite{Freiman91,Brodyanskii93,Freiman09,Freiman05,Hetenyi05}.  Moreover, as models
of coupled rotors are used to describe other physical phenomena
(granular superconductors~\cite{Parmenter67,Simanek85,Simkin91}) the behaviour of
the model itself is of interest.

\subsection{The cases of infinite and zero spin conversion time}

There are two common approaches to systems with coupled rotors
corresponding to homonuclear molecules with nuclear spin coupled to
angular momenta.  The zero relaxation time limit corresponds to
averaging the partition functions over the spin configurations.  The
one-particle partition function can be written in this case as
\begin{equation}
Q_{\mbox{tot}} = \sum_i G_i Q_i,
\label{eqn:Q_0}
\end{equation}
where $G_i$ denotes the degeneracy of angular momentum states
corresponding to a particular spin configuration, and $Q_i$ denotes
the partition function in which the states which enter are the ones
that are coupled to the same spin configuration, i.e.
\begin{equation}
  Q_i = \mbox{Tr}_i \{\mbox{exp}(-\beta H_{MF})\},
\end{equation}
where $\mbox{Tr}_i$ denotes tracing over a particular manifold of
angular momentum states.  In
this case the distribution of states depends on the parameters, i.e.
temperature, coupling constant.  Minimizing the free energy
corresponding to Eq. (\ref{eqn:Q_0}) results in 
\begin{equation}
\Gamma = \langle \mbox{cos} (n \phi) \rangle,
\end{equation}
and the fraction of a particular species at thermal equilibrium is given by
\begin{equation}
X_i = \frac{G_i Q_i}{\sum_j G_j Q_j}.
\end{equation}
The fraction of a particular species will in general be a function of
the parameters defining the system (temperature, coupling and
rotational constant).

The infinite relaxation time limit in the mean-field approximation
corresponds to writing the partition function as a product
\begin{equation}
  Q_{\mbox{tot}} = \prod_i Q_i^{X_i},
\label{eqn:Q_tot_inf}
\end{equation}
with the fraction of species $i$ defined as $X_i=N_i/N$.  Here $Q_i$
again denotes a partition function into which the angular momentum
states that enter are the ones corresponding to a particular spin
configuration.  This way of writing the partition function is only
possible in mean-field theory, where the many-body Hamiltonian
separates into a sum of additive single-rotor Hamiltonians.  From Eq.
(\ref{eqn:Q_tot_inf}) it follows that the expression for the free
energy is
\begin{equation}
F_{\mbox{tot}} = \sum_i X_i F_i,
\end{equation}
where $F_i$ denote the free energies of molecules belonging to a
particular manifold of angular momentum states.  The free energy is
minimized if
\begin{equation}
  \Gamma = \sum_i X_i \Gamma_i,
\end{equation}
where
\begin{equation}
\Gamma_i = \langle \mbox{cos}(n\phi)\rangle_i,
\label{eqn:mfinf}
\end{equation}
where $\langle \rangle_i$ denotes averaging over a particular manifold
of angular momentum states.

\subsection{The case of slow but finite spin conversion}
\label{ssec:slow}

In the following the mean-field theory of systems where spin
conversion is slow but finite will be considered.  The equilibration
of the rotors is instantaneous, but the system finds its equilibrium
proportion of different spin configurations on a longer time-scale.
The starting point is Eq. (\ref{eqn:Q_tot_inf}) but the issue is the
optimal proportion of different spin configurations (i.e. $N_i$).  The
probability of a particular set of $N_i$s is given by
\begin{equation}
P(\{N_i\}) = \frac{C(\{N_i\}) \prod_i Q_i^{N_i}}{Z},
\label{eqn:P}
\end{equation}
i.e. the partition function of Eq. (\ref{eqn:Q_tot_inf}) multiplied by
$C(\{N_i\}$ the number of ways such a configuration can occur.  $Z$ is
the normalization constant
\begin{equation}
Z = \sum_{\{N_i\}} P(\{N_i\}).
\end{equation}
$C(\{N_i\}$ is a combinatorial factor which can be written as
\begin{equation}
C(\{N_i\} = \frac{N!}{\prod_i{\left[\left(\frac{N_i}{G_i}\right)\Large{!}\right]^{G_i}}}.
\label{eqn:C}
\end{equation}
Note that here the number of rotors are taken to be evenly distributed
between the different degenerate nuclear spin states within an angular
momentum manifold.  Such a distribution corresponds to maximizing the
entropy, and in the thermodynamic limit it can be expected that one
configuration dominates.  In the absence of fields which break the degeneracy
this assumption can be expected to be correct.

In the thermodynamic limit Stirling's formula can be applied to the
combinatorial factor $C(\{N_i\})$ and obtain
\begin{equation}
C(\{N_i\}) = \mbox{exp}\left[-N\left\{\sum_i X_i\mbox{ln}(X_i/G_i)\right\}\right],
\end{equation}
where $X_i=N_i/N$.  Hence the overall probability of a particular
configuration becomes
\begin{equation}
P(\{N_i\}) = \mbox{exp}\left[-N\left\{\sum_i X_i\mbox{ln}\left(\frac{X_i}{G_iQ_i}\right)\right\}\right].
\label{eqn:prob}
\end{equation}
Clearly, in the thermodynamic limit, one set of values of $\{X_i\}$
will dominate, so the exponent of Eq.  (\ref{eqn:prob}) can be
optimized under the normalization constraint.  The resulting condition
is
\begin{equation}
\mbox{ln}\left(\frac{X_i}{G_iQ_i}\right) + \left(1-\frac{X_i}{Q_i}\frac{\partial Q_i}{\partial X_i}\right) + \lambda = 0,
\label{eqn:cond}
\end{equation}
where $\lambda$ denotes the Lagrange multiplier resulting from the
normalization.  The second $Q_i$ dependent term in Eq.
(\ref{eqn:cond}) can be evaluated as
\begin{equation}
\frac{\partial \mbox{ln}Q_i^{X_i}}{\partial X_i} = \beta \Gamma_i (\langle \mbox{cos}(n\phi)\rangle_i -\Gamma_i),
\end{equation}
which evaluates to zero as the $\Gamma_i$'s in this case are also fixed by the mean-field condition (Eq. (\ref{eqn:mfinf})).  Hence Eq. (\ref{eqn:cond}) gives
\begin{equation}
X_i = \frac{G_i Q_i}{\sum_j G_j Q_j},
\label{eqn:X_i}
\end{equation}
which is the distribition in the zero relaxation time limit.
Substituting the expression for $X_i$ from Eq. (\ref{eqn:X_i}) into Eq
(\ref{eqn:prob}) gives the partition function for the zero spin
conversion limit as expected.  

An interesting result of the above analysis is that the entropy of the
system in which spin conversion times are long but allowed to
equilibrate can be broken up into two pieces: a configurational
entropy arising from the combinatorical prefactor counting the
configurations $C(\{N_i\})$ and one arising from the product of single
particle partition functions which here will be called the rotational
entropy.  The total entropy per molecule calculated from Eq.
(\ref{eqn:prob}) can be written
\begin{equation}
S = S_{\mbox{conf}} + S_{\mbox{rot}},
\label{eqn:S_tot}
\end{equation}
where
\begin{equation}
S_{\mbox{conf}} = - \sum_i X_i \ln \left(\frac{X_i}{G_i} \right)
\label{eqn:S_conf}
\end{equation}
and
\begin{equation}
S_{\mbox{rot}} = \sum_i X_i S_i,
\label{eqn:S_rot}
\end{equation}
with
\begin{equation}
S_i = \ln Q_i - \beta \frac{\partial \ln Q_i}{\partial \beta}.
\label{eqn:S_i}
\end{equation}
Eq. (\ref{eqn:S_i}) is a weighted sum over the entropy per molecule
associated with the rotational states coupled to a particular spin
configuration.

\section{Nuclear spin in three-fold symmetric uniaxial rotors}
\label{sec:nspin}

The model studied here is one of three-fold symmetric rotors
performing uniaxial rotation.  The Hamiltonian is of the form given in
Eq. (\ref{eqn:Hml}), with $n=3$.  As this model can correspond to
compounds containing ammonia or methyl groups, where identical atoms
are performing the rotation, nuclear spin can be expected to influence
properties such as orientational ordering.

The issue of nuclear spin in the case of uniaxial three-fold symmetric
rotors is qualitatively different from that in the two-fold symmetric
case.  In the latter it is the fermionic or bosonic character of the
constituent atoms that determine which nuclear spin functions
couple to odd or even angular momenta.  The effects of nuclear spin in
the three-fold symmetric case follow from the fact that a rotation of
$\frac{2\pi}{3}$ or $-\frac{2\pi}{3}$ correspond to two permutation
exchanges of nuclei, hence no change in the overall wavefunction
occurs either for fermions or bosons~\cite{Press81}.  

For free uniaxial rotors wavefunctions which remain unaltered when
rotated by $\frac{2\pi}{3}$ can be constructed as follows.  The
molecular wavefunction can be written as a product,
\begin{equation}
\Psi_{s,m_s}(\phi) = \frac{1}{\sqrt{2\pi}}\mbox{exp}(i m \phi)\Sigma_{s,m_s}
\end{equation}
where $\phi$ indicates the coordinate of the rotor, and $s,m_s$ indicate the
spin quantum numbers of the three nuclei in the coupled angular momentum
representation, and $\Sigma_{s,m_s}$ denotes the spin part of the
wavefunction.  Rotating by $\frac{2\pi}{3}$ gives
\begin{eqnarray}
\hat{R}\left(\frac{2\pi}{3}\right)\Psi_{s,m_s}(\phi) = \frac{1}{\sqrt{2\pi}}
\mbox{exp}\left\{i m \left(\phi+ \frac{2\pi}{3}\right)\right\} \\
\times \hat{R}\left(\frac{2\pi}{3}\right)\Sigma_{s,m_s}. \nonumber
\end{eqnarray}
It can be shown~\cite{Press81} that the spin eigenfunctions in the
coupled representation are eigenfunctions of the rotation operator
$\hat{R}\left(\frac{2\pi}{3}\right)$ with possible eigenvalues $1,
e^{i\frac{2\pi}{3}},e^{-i\frac{2\pi}{3}}$.  It follows that a given
spin-eigenfunction can have only certain angular momentum states, so
that the requirement that a rotation by $\frac{2\pi}{3}$ causes no
change in the wavefunction is satisfied.  In particular spin
eigenfunctions with eigenvalues
$1,e^{i\frac{2\pi}{3}},e^{-i\frac{2\pi}{3}}$ couple to angular
momentum states $m=3n,3n-1,3n+1$ respectively, where $n$ is an
integer.

The ammonia molecules in inclusion compounds consist of nearly freely
rotating NH$_3$ and ND$_3$ molecules.  In the case of an NH$_3$ group,
where the rotating atoms are of spin $\frac{1}{2}$, it can be shown
that the $s=\frac{3}{2}$ states are $m=3n$, whereas one group of the
$s=\frac{1}{2}$ states states are of the $m=3n-1$ the other of the
$=3n+1$ variety.  In ND$_3$ the $s=3$, $s=0$, and one of the three
$s=1$ states are $m=3n$, one of the $s=2$ and $s=1$ states are
$m=3n-1$, the remaining $s=2$ and $s=1$ being $m=3n+1$.  In group
theoretical terms the $m=3n$ states form the one-dimensional
representation $A$ of the group $C_3$, the states $m=3n+1$ and
$m=3n-1$ form the representations $E_a$ and $E_b$.  In the following
the rotating atoms are assumed to be of spin-$\frac{1}{2}$.

\begin{figure}[htp]
\vspace{1cm}
\psfig{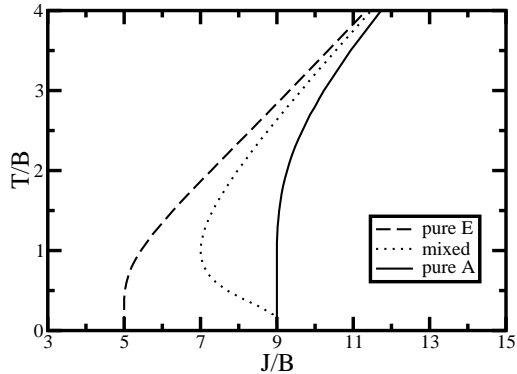}
\vspace{1cm}
\caption{Phase diagrams calculated via mean-field theory.  Systems
  with pure $A$ and pure $E$ species, and with the thermal equilibrium
  distribution are shown.  The thermal equilibrium phase diagram
  corresponds to the zero spin conversion time limit, as well as the
  case of thermally equilibrated rotors with long spin conversion times on
  the time scale of rotations.}
\label{fig:pdct0}
\end{figure}

\begin{figure}[htp]
\vspace{1cm}
\psfig{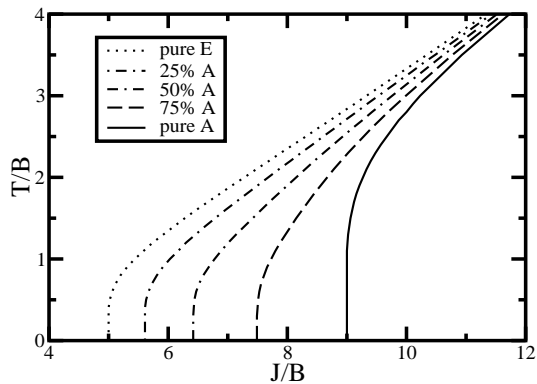}
\vspace{1cm}
\caption{Phase diagrams calculated via mean-field theory.  Comparisons
  are shown for different proportions of $A$ and $E$ species.  All
  phase diagrams calculated in the infinite spin conversion time
  approximation.  The tendency to order increases with the proportion of
  $E$ molecules.}
\label{fig:pdctinf}
\end{figure}

\begin{figure}[htp]
\vspace{1cm}
\psfig{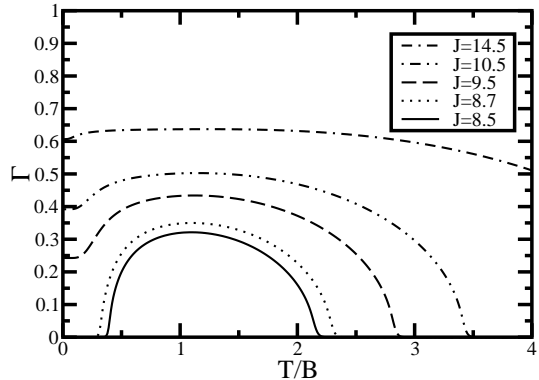}
\vspace{1cm}
\caption{Order parameter as a function of temperature for
  different coupling constants.  The coupling constants $J=8.5$ and
  $J=8.7$ correspond to the reentrant regime of the phase diagram.
  $J=9.5$ and $J=10.5$ is in the ordered regime, but shows
  qualitatively similar behaviour to the reentant ordering, the order
  parameter increases at low temperatures.  This tendency is
  suppressed in the classical limit ($J=14.5$).}
\label{fig:op}
\end{figure}

\begin{figure}[htp]
\vspace{1cm}
\psfig{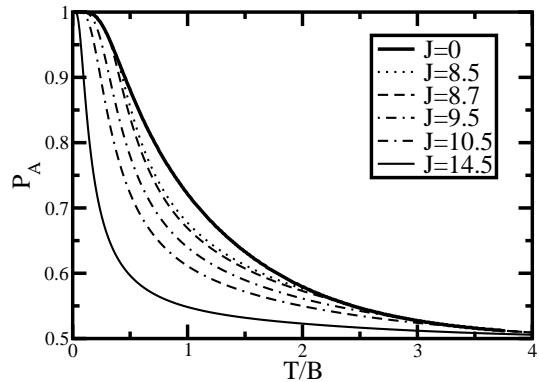}
\vspace{1cm}
\caption{Proportion of $A$ species as a function of temperature for
  different coupling constants.  For $J=8.5$ and $J=8.7$ the regimes
  of finite order parameter (Fig. \ref{fig:op}) correspond to where
  the distribution deviates from the $J=0$ case.  For the couplings
  $J=9.5$ and $J=10.5$ the increase in order parameter as a function
  of temperature coincides with where the distribution function
  deviates from the $J=0$ case.  Hence ordering is correlated with an
  increase in the proportion of $E$ species.}
\label{fig:xa}
\end{figure}

In Refs.  \onlinecite{Hetenyi05,Hetenyi05b} an extension of mean-field
theory was developed in which it was assumed that each lattice site is
of a particular spin state, but the rotational degrees of freedom were
solved using coupled mean-field equations.  This method was used to
study solid hydrogen.  The ratio of ortho versus para species was
approximated as that of the free rotor system and the phase boundary
was determined.  The result that the mean-field theory of equilibrated
spin conversion in the case of long spin conversion times is
equivalent to the case when spin conversion times are instantaneous
lends support to the validity of the approximation used in Refs.
\onlinecite{Hetenyi05,Hetenyi05b}.

\section{Results}
\label{sec:results}

In Figs. \ref{fig:pdct0} and \ref{fig:pdctinf} the mean-field phase diagrams
are shown.  The thermally equilibrated species shows a reentrant phase
diagram (a result also found previously Freiman and Tretyak~\cite{private}).
This phenomenon is common in models of coupled
rotors~\cite{Freiman91,Brodyanskii93,Freiman09,Shin09,Hetenyi05,Freiman05,Mueser99,Hetenyi99,Hetenyi05b,Simanek85,Simkin91},
and is caused by the fact that the rotational state that lies higher in
energy has a stronger ordering tendency.  This idea is corroborated by the
results in figure \ref{fig:pdct0}: the pure $E$ system orders at a lower
coupling constant than the pure $A$ system, hence it has a stronger tendency
to order.  As the temperature of the thermally mixed system increases more
rotors can access $E$ states hence the system orders.  The results for the
infinite spin conversion time limit (\ref{fig:pdctinf}) show progressively
stronger tendency to order as the proportion of $E$ species is increased.
Note that the behaviour found here for three-fold symmetric rotors is
qualitatively different from hydrogen, where the ortho species has an ordered
ground state for finite coupling constant, and for any small fraction of the
ortho species~\cite{Hetenyi05,Freiman05,Hetenyi05b}.

In Fig. \ref{fig:op} the order parameter $\Gamma$ is shown as a
function of temperature at different coupling constants.  In the
reentrant regime ($J=8.5$ and $J=8.7$) the two phase transitions, the
high temperature one caused by thermal excitations, and the low
temperature one caused by quantum fluctuations are clearly visible.
For higher values of the coupling constant $J$ ($J>9.5$) one still
sees increased ordering as the temperature is increased from zero.
This tendency is suppressed as the classical limit is approached.  The
ordering tendency can be understood in all cases in terms of the
distribution of the various spin nuclear states shown in Fig.
\ref{fig:xa}.  The thick line shows the distribution in the $J=0$
case, where the system is disordered.  In the regime where quantum
melting occurs the deviation of the curves corresponding to $J=8.5$
and $J=8.7$ coincide with the range of nonzero order parameter.  For
higher $J$ values the point where the order parameter begins to
increase as a function of temperature corresponds to where the
distribution functions begins to deviate from the free rotor case.
\begin{figure}[htp]
\vspace{1cm}
\psfig{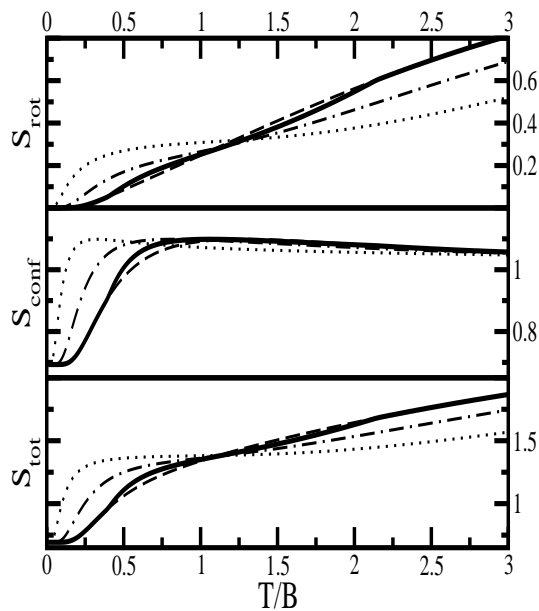}
\vspace{1cm}
\caption{Rotational, configurational, and total entropy for a thermally
  equilibrated system at $J=8.5$.  The thick line shows the results
  with the order parameter minimized.  The other curves show results
  at fixed order parameter: dashed line $\Gamma=0.0$, dot-dashed line
  $\Gamma=0.5$, dotted line $\Gamma=1.0$.  All three graphs show that
  the ordered state has higher entropy at low temperature than the
  disordered state, and this situation reverses at higher
  temperatures.  The configurational entropy (middle panel) shows a
  shift in the maximum to the left as the ordering is increased.}
\label{fig:s}
\end{figure}

It has been argued that the reentrance is driven by anomalous
behaviour of the entropy\cite{Freiman05}.  Freiman {\it et al.} have
shown that if the order parameter is held fixed the entropy is
higher in the ordered state at low temperatures, and the situation
reverses as the temperature increases.  In Fig. \ref{fig:s} the
entropy and its components (rotational and configurational, see Eq.
(\ref{eqn:S_tot})) are shown for fixed order parameters, as well as
the entropy of the thermally equilibrated solution for $J=8.5$.  Both
the rotational and configurational entropies show the entropy anomaly
described by Freiman {\it et al.}.  However the behaviour of the
configurational and rotational entropies is qualitatively different.
The configurational entropy displays a maximum which shifts to smaller
values of temperature for the states which are more ordered.  At large
temperature the configurational entropies converge.  For the
configurational entropy the tendency of the entropy to increase with
order at low temperatures is very pronounced, whereas at higher
temperatures, the opposite tendency is significantly weaker.  From
Fig. \ref{fig:s} one can argue that the low temperature reentrance
anomaly is driven by both the configurational and rotational
entropies.

\section{Conclusions}
\label{sec:conclusions}

In this study the mean-field theory of coupled quantum rotors was
investigated.  Particular attention was devoted to rotor systems which
correspond to molecules where identical atoms rotate.  In such systems
the nuclear spin states couple to angular momentum states, and thereby
effect orientational ordering.  As the spin conversion times
in such systems tend to be slow, experiments can be performed at fixed
ratios of different spin states or at thermal equilibrium.

The mean-field approach to such systems is usually based on two strategies:
in one case the spin conversion is assumed to be instantaneous on the time
scale of rotation, in the other infinite.  In the former case all angular
momentum states are allowed for each molecule.  In the latter a particular
molecule is allowed only one manifold of angular momentum states fixed by the
symmetry of the overall wavefunction.  In this study the case of long but
finite spin conversion times was considered, i.e., when spin conversion times
greatly exceed rotational time-scales, but the experiment is carried out such
that the distribution of molecules in different nuclear spin states is
allowed to equilibrate.  It was shown that in the mean-field approximation
such an approach yields the same results as the instantaneous spin conversion
time limit, provided that the system is in the thermodynamic limit.  Another
outcome of the formalism is that the entropy becomes a sum of a
configurational and a rotational contribution.

Calculations were also performed to determine the phase diagram of coupled
threefold symmetric rotors.  In this case the nuclear spin causes each
molecule to be either of three types, $A$, $E_a$, and $E_b$, of which $E_a$
and $E_b$ are energetically equivalent and show the same ordering, hence $A$
and $E$.  Pure $A$ systems show a weaker tendency to order than pure $E$
systems.  Zero spin conversion time and thermal equilibration leads to a
reentrant phase transition confirming previous results~\cite{private}.  It
was also shown that the entropy anomaly known to exist in reentrant
systems~\cite{Freiman05} is present in both the configurational and the
rotational contributions.  At low temperatures the ordered state has higher
entropy than the disordered state in both cases.  The configurational entropy
displays a maximum which shifts to lower temperatures with increasing order.

\begin{acknowledgements}
Part of this work was performed at the Institut f\"ur Theoretische
Physik at TU-Graz under FWF (F\"orderung der wissenschaftlichen
Forschung) grant number P21240-N16.  Part of this work was performed
under the HPC-EUROPA2 project (project number 228398).  Helpful
discussions with Y. A.  Freiman and S. Scandolo are gratefully
acknowledged.
\end{acknowledgements}



\begin{references}

\bibitem{Silvera80} I.~F. Silvera, {\it Rev. Mod. Phys.} {\bf 52} 393 (1980).

\bibitem{Mao94} H.~K. Mao and R.~J. Hemley, {\it Rev. Mod. Phys.} {\bf
    66} 671 (1994).

\bibitem{deg} The high(low)spin degeneracy state is also known as
  ortho(para).

\bibitem{Cui95} L.~J. Cui, N.~H. Chen, and I.~F. Silvera, {\it Phys.
    Rev. B} {\bf 51} 14997 (1995).

\bibitem{Freiman91} Yu. A. Freiman, V.~V. Sumarokov, A.~P.
  Brodyanskii, and A. Jezowski, { \it J. Phys.: Cond. Matter}, {\bf
    3} 3855  (1991).

\bibitem{Brodyanskii93} A.~P.  Brodyanskii, V.~V. Sumarokov, Yu. A. Freiman,
  and A. Jezowski, { \it Low Temp. Phys.}, {\bf 19} 368 (1993).

\bibitem{Freiman09} Y.~A. Freiman, B. Het\'enyi, and S.~M. Tretyak, in {\it
    Metastable Systems Under Pressure, (NATO Science for Peace and Security
    Series A: Chemistry and Biology)}, Eds. S.~J. Rzoska, A. Drozd-Rzoska,
  V. Mazur, Springer-Verlag (2009).  See also:arXiv:0907.1299.

\bibitem{Moshary96} F. Moshary, N.~H. Chen, and I.~F. Silvera, {\it Phys. Rev.
    Lett.}, {\bf 71} 3814 (1996).

\bibitem{Shin09} H. Shin and Y. Kwon, {\it J. Korean Phys. Soc.}, {\bf 54}
  1582 (2009).

\bibitem{Hetenyi05} B. Het\'enyi, S. Scandolo, and E. Tosatti, {\it
    Phys. Rev. Lett.}, {\bf 94} 125503 (2005).

\bibitem{Freiman05} Yu. A. Freiman, S.~M. Tretyak, H.-K. Mao, and
  R.~J. Hemley, { \it J. Low Temp. Phys.}, {\bf 139} 765 (2005).

\bibitem{Goncharenko05} I. Goncharenko and P. Loubeyre, {\it Nature}, {\bf 435}
  1206 (2005).

\bibitem{Mueser99} M.~H. M\"user and J. Ankerhold, {\it Europhys.
    Lett.}, {\bf 44} 216 (1999).

\bibitem{Hetenyi99} B. Het\'enyi, M.~H. M\"user, and B.~J. Berne, {\it Phys.
   Rev. Lett.}, {\bf 83} 4606 (1999).

\bibitem{Hetenyi05b} B. Het\'enyi, S. Scandolo, and E. Tosatti, {\it
    J. Low Temp. Phys.}, {\bf 139} 753 (2005).

\bibitem{Martonak97} R. Marto\v{n}\'ak, D. Marx, and P. Nielaba, {\it Phys.
    Rev. E}, {\bf 55} 2184 (1997).

\bibitem{Parmenter67} R.~H. Parmenter, {\it Phys. Rev.}, {\bf 154} 353
  (1967).

\bibitem{Simanek85} E. \v{S}im\'{a}nek, {\it Phys. Rev. B}, {\bf 32} 500 (1985).

\bibitem{Simkin91} M.~V. Simkin, {\it Phys. Rev. B}, {\bf 44} 7074 (1991).

\bibitem{Sloan90} E.~D. Sloan, {\it Clathrate Hydrates of Natural
    Gases} (Marcel Bekker, New York, 1990).

\bibitem{Hubbard81} W.~B. Hubbard, {\it Science}, {\bf 214} 145 (2001).

\bibitem{Lunine87} J.~I. Lunine and D.~J. Stephenson, {\it Icarus}, {\bf 70} 61 (1987).

\bibitem{Loveday01} J.~S. Loveday, R.~J. Nelmes, M. Guthrie, S.~A. Belmonte, D.~R. Allan, D.~D. Klug, J.~S. Tse, and Y.~P. Handa {\it
    Nature}, {\bf 410} 661 (2001).

\bibitem{Buettner94} H.~G. B\"uttner, G.~J. Kearley, S.~J. Howard, and
  F. Fillaux, { Acta Crystallogr. B} {\bf 50} 431 (1994).

\bibitem{Iwamoto74} T. Iwamoto, T. Miyoshi, and Y. Saaki, { Acta Crystallogr. B} {\bf 30} 292 (1974).

\bibitem{Sobolev04} O. Sobolev, P. Vorderwisch, and A. Desmedt, {\it
    Chem. Phys.}, {\bf 308} 147 (2004).

\bibitem{Press81} W. Press, {\it Single-Particle Rotations in
    Molecular Crystals}, (Springer Tracts in Modern Physics, 1981).

\bibitem{Prager97} M. Prager and A. Heidemann, {\it Chem. Rev.} {\bf
    97} 2933 (1997).

\bibitem{Wurger90} A. W\"urger, {\it Z. Phys. B}, {\bf 81} 273 (1990).

\bibitem{private} Y.A. Freiman and S. M. Tretyak, private communication.

\end{references}
\end{document}